\documentclass[preprint]{aastex}

\slugcomment{Accepted for publication by AJ}

\shorttitle{Dust Opacities in Spiral Galaxy Disks}
\shortauthors{Holwerda et al.}

\begin{document}

\title{The Opacity of Spiral Galaxy Disks III: \\ Automating the ``Synthetic Field Method''}

\author{B. W. Holwerda\altaffilmark{1,2}, R. A. Gonzalez \altaffilmark{3} and Ronald J. Allen\altaffilmark{1}}

\email{holwerda@stsci.edu}
\and
\author{P. C. van der Kruit \altaffilmark{2}}

\altaffiltext{1}{Space Telescope Science Institute, Baltimore, MD 21218,USA}
\altaffiltext{2}{Kapteyn Institute, Groningen Landleven 12, 9747 AD Groningen, The Netherlands}
\altaffiltext{3}{Centro de Radiastronom\'{\i}a y Astrof\'{\i}sica, Universidad Nacional Aut\'{o}noma de M\'{e}xico, 58190 Morelia, Michoac\'{a}n, Mexico}

\clearpage
\newpage

\begin{abstract}
The dust extinction in spiral disks can be estimated from the counts of background field galaxies, provided the deleterious effects of confusion introduced by structure in the image of the foreground spiral disk can be calibrated. \cite{Gonzalez98} developed a method for this calibration, the ``Synthetic Field Method'' (SFM), and applied this concept to a HST/WFPC2 image of NGC4536.
The SFM estimates the total extinction through the disk without the necessity of assumptions about the   distribution of absorbers or the disk light. The poor statistics, however, result in a large error in individual measurements.
We report on improvements to and automation of the Synthetic Field Method which render it 
suitable for application to large archival datasets. To illustrate the strengths and weaknesses of this new method, the results on NGC 1365, a SBb, and NGC 4536, a SABbc, are presented. 
The extinction estimate for NGC1365 is $A_I = 0.6_{-0.7}^{+0.6}$ at 0.45 $R_{25}$ and for NGC4536 it is $A_I = 1.6_{-1.3}^{+1.0}$ at 0.75 $R_{25}$. 
The results for NGC4536  are compared with those of \cite{Gonzalez98}. The automation is found to limit the maximum depth to which field galaxies can be found. Taking this into account, our results agree with those of \cite{Gonzalez98}. 
We conclude that this method can only give an inaccurate measure of extinction for a field covering a  small solid angle. An improved measurement of disk extinction can be done by averaging the results over a series of HST fields, thereby improving the statistics. This can be achieved with the automated method, trading some completeness limit for speed.
The results from this set of fields are reported in a companion paper \citep{Holwerda04b}.
\end{abstract}

       
\section{Introduction}

The question of how much the dust in spiral galaxies affects our perception of them became a controversial topic after \cite{Valentijn90} claimed that spiral disks were opaque. Valentijn based his conclusion on the apparent independence of disk surface brightness on inclination. \cite{Disney90} objected to this conclusion, claiming instead that galaxy disks are virtually transparent and that Valentijn's results were due to a selection effect.  Others joined the controversy and within a few years a conference was organized to address the question of how best to determine galaxy disk opacity and what results could be obtained \citep{Cardith94}.

Notably, \cite{kw92} proposed a method to determine the opacity of a foreground disk galaxy in the rare cases where it partially occults another large galaxy.
This technique has been followed up extensively with ground-based optical and infrared imaging  \citep{Andredakis92,Berlind97,kw99a,kw00a}, spectroscopy  \citep{kw00b}, and HST imaging  \citep{kw01a,kw01b,Elmegreen01}.
Their results indicated higher extinction in the arms and a radial decrease of extinction in the inter-arm regions. Also the highest dust extinction was found in the areas of high surface brightness.  Their sample of $\sim$20 suitable pairs seems now to be exhausted. This method furthermore assumes symmetry for the light distribution of both galaxies, so that a method independent of the light distribution is needed to confirm these results.

Instead of a single large background galaxy, the general field of distant galaxies can be used as a background source. \cite{Hubble34} noted the apparent reduction in the surface density of background galaxies at lower Galactic latitudes. \cite{Burstein82} published a map of Galactic extinction based on the galaxy counts by \cite{Shane67}. 
Studies of extinction in the Magellanic clouds based on field galaxy counts have been done regularly \citep{Shapley51,Wesselink61a,Hodge74,MacGillivray75,Gurwell90,Dutra01}.  
More recently, attempts have been made to use field galaxy counts in order to establish the opacity in specific regions of nearby foreground galaxies from ground based data \citep{Zaritsky94,Lequeux95,Cuillandre01}. Occasionally, the presence of field glaxies in a spiral disk is presented as anecdotal evidence of galaxy transparency \citep{Roennback97, Jablonka98}. The results of these studies have suffered from the inability to distinguish real opacity from foreground confusion as the reason for the decrease in field galaxy numbers. 

\cite{Gonzalez98} (Paper I) introduced a new approach to calibrate foreground confusion which they called the ``Synthetic Field Method''. 
While this new method can provide the required calibration for specific foreground galaxies, it is labor intensive and therefore ill-suited for the study of larger samples of galaxies of various types.
In this paper we present the first results from a project to automate major steps in the Synthetic Field Method. After first providing a brief summary of the major features of the method, we describe how it was automated and which improvements we have made. As an illustration we have applied the new automated method to two galaxies, NGC 4536 and NGC 1365, and compared the results of our improved algorithms to those obtained on the former galaxy by \cite{Gonzalez98}. 
In a companion paper \citep{Holwerda04b} we report on our application of the method to a data set consisting of 32 HST/WFPC2 pointings on 29 nearby galaxies.

\section{The Synthetic Field Method}

Figure 1 shows a schematic of how the Synthetic Field Method is applied. Deep exposures of a nearby galaxy are obtained with the Wide Field Planetary Camera 2 (WFPC2) on the Hubble Space Telescope and background field galaxies are identified. Synthetic fields are then created by adding exposures from the Hubble Deep Fields \citep{HDFNobs,HDFSobs} and the background field galaxy counts are repeated.  The ratio of the surface density of real field galaxies to that of the HDF galaxies for any given region is a measure of the opacity in that region of the foreground galaxy. In practice, a series of synthetic fields is created with successively larger extinctions applied to the HDF galaxies until a match is obtained with the real field galaxy count, thus providing a quantitative measure of opacity in the foreground system.
The method provides a way of calibrating the deleterious effects of confusion caused by the granular structure of dust, stars and luminous gas in the foreground galaxy. These effects are dramatic; for example a typical single WFPC2 chip (1\farcm2 $\times$ 1\farcm2) in the HDF may contain some 120 easily-identified background galaxies. This can drop to only 20 or 30  HDF galaxies when the foreground galaxy is present, and this number becomes even smaller for the larger values of a simulated foreground opacity. Counting background galaxies is therefore a battle against small number statistics, and reliable results are difficult to obtain on a single foreground galaxy. However, the method can in principle be applied to many nearby galaxies, and average opacities obtained e.g. as a function of radius for a sample of galaxies of similar morphological type.

\subsection{Limits of the Synthetic Field Method}

 \cite{Gonzalez03} (Paper II) have discussed the broad limitations of the method in terms of the optimum distance interval for which it can be used most effectively given current and future ground and space-based imaging instruments. 
Two effects compete to limit the distance to which the SFM can be applied:
First, if the foreground galaxy is too close, confusion from the ``granularity'' in the images, caused by a more resolved foreground disk, further reduces the number of bona fide field galaxies. Second, if the foreground galaxy is too distant, the small area of sky covered also reduces the numbers of field galaxies that can be used. These two effects conspire such that the optimum distance range for the Synthetic Field Method with HST/WFPC2 observations is approximately between 5 and 25 Mpc. Within this range there are of course many hundreds of galaxies visible to HST. However, now we are hindered by another limitation of the specific implementation of the Synthetic Field Method used by \cite{Gonzalez98}, namely, that the identification of background and synthetic galaxies was carried out entirely visually, a time-consuming and laborious
process. Clearly if any real progress is to be made, the process of identifying the field galaxies has to be automated \citep{Holwerda01b, Holwerda02dunk}. Automated field galaxy identification has the benefits of speed and consistency across the datasets. Conversely in paragraph 5.1 we shall illustrate how it imposes a brightness limit on the selected objects.

\section{Automation of the method}

We have automated three steps in the Synthetic Field Method: first, the processing of archived exposures to produce combined images; second, the construction of catalogs of objects of simulated and science fields; and finally, the automatic selection of candidates for field galaxies, based on the parameters in the catalogs. A visual control on the process was retained by reviewing the final list of candidate field galaxies in the science field. We will now describe each step in the entire process in more detail.

\subsection{Processing archival WFPC2 data}

When data sets are recovered from the HST archive, the most recent
corrections for hot pixels, bad columns, geometric
distortions and the relative Wide Field (WF) and the Planetary Camera (Pc) CCD positions for the observation date are applied in
the archive's pipeline reduction. This pipeline system provides the user with a science
data file and a quality file with positions of the bad pixels
\citep{1999adass...8..203S}.

In order to stack multiple exposures, corrected for small position 
shifts and with the cosmic rays removed, we used the 'drizzle'-method \citep{Fruchter97a}, 
packaged in routines under IRAF/{\it pyraf}, with an output pixel scale of 0.5 of the original pixel 
and PIXFRAC between 0.8 and 1.0, depending on the number of shifts in the 
retrieved data. 
The PIXFRAC parameter sets the amount by which the input pixel is
shrunk before it is mapped onto the output plane; a PIXFRAC lower 
then unity improves the sampling of the stacked image. 

We developed a custom script to combine all exposures using {\it python} with the {\it pyraf}
package \citep{pyraf1}, based on examples in the Dither Handbook
\citep{drizzlehandbook}. 
The images are prepared for crosscorrelation in order to find the
relative shifts; the background was subtracted ({\it sky}) and all
none-object-pixels were set to zero ({\it precor}).
Subsequently crosscorrelation images between exposures were made for
each of the four CCDs ({\it crosscor}). The fitted shifts from these
were averaged ({\it shiftfind, avshift}). 
Any rotation of an exposure was calculated from the header information 
and ultimately derived from the spacecraft orientation provided by the guide 
stars.\footnote{The uncertainty in the orientation angle is the result of uncertainties of a few arcseconds in the positions of guide stars with a separation of a few arcminutes. Therefore, this uncertainty is an order of magnitude smaller than the uncertainties in right ascension and declination of the pointing.}
All original exposures were shifted to the reference coordinates and a median 
image was constructed from these ({\it imcombine}). The median image was 
then copied back to the original coordinates ({\it blot}). The cosmic rays in each
exposure were identified from the difference between the shifted median
image and the original exposure ({\it driz\_cr}). A mask with the
positions of the cosmic rays and hot pixels, identified in the data quality
file, was made for each exposure.

The exposures were drizzled onto new images and separately onto a mosaic with
cosmic rays and bad pixels masked off ({\it drizzle,loop\_gprep}). The new pixel-scale
is fixed at 0\farcs05 but, to check the choice of PIXFRAC value, the script computed 
the rms of the weight image output from drizzle. The rms standard deviation should be between 15 and 30\% of the mean.
 
For both galaxies used as examples here, there are many exposures made over several epochs. However the shifts for NGC 4536 are smaller than one original pixel (0\farcs1), and the bad columns of the CCD detector are unfortunately not covered by good pixels from other exposures (see Figure 2). 
Several exposures for NGC 1365 display shifts greater then a pixel, which helps to cover the bad columns and results in a cleaner looking image (Figure 8).
In both cases the number of shifts was sufficient for a PIXFRAC of 0.8 with the new pixel scale of 0\farcs05.

\subsection{Making object catalogs}

A modified version of Source Extractor v2.2.2 (hereafter SE, \cite{SE}) was used
to generate catalogs of objects for the science fields and simulations. The
F814W (``I''-band) fields were used for detection. Catalogs for the F555W (``V''-band) fields
were constructed using the dual mode; the photometry was done on the V field using the
I apertures. All the structural parameters were derived from the I
images. In Table 1 we list our choice of SE input settings. Table 2
lists the intrinsic output parameters from SE and Table 3 the new output parameters we added. In addition the position of objects on the CCD and on the sky are in the catalogs.

It was already noted by \cite{SE} that the success of SE's native star/galaxy
classification parameter was limited to the very brightest objects. Several
other parameters are described in the literature for the classification of field
galaxies. \cite{Abraham94} and \cite{Abraham97}
used asymmetry, contrast and concentration to identify Hubble type of
galaxies. Similarly \cite{Conscelice97, Conscelice99, Conscelice00a, Conscelice00b, CAS} used
asymmetry, concentration and clumpiness as classifiers. By adding some
of these parameters or our approximations of them to the Source Extractor 
code, we obtained a better parameter space within which to
separate field galaxies from objects in the foreground
galaxy. 

\subsection{Selection of field galaxy candidates using ``fuzzy boundaries''}

The characteristics as determined by SE for field galaxies and
foreground objects are very similar; for example, we show in Figure 3 the distribution 
of the FWHM of all objects and that of the HDF galaxies. This similarity exists because there
are many extended foreground objects: star clusters, HII regions, artifacts
from dust lanes, diffraction spikes near bright stars and ``objects''
which are blends of several objects. The field galaxies also span a range in
characteristics, as can be seen in the Hubble Deep Fields. Simple cuts 
in parameter space can do away with some objects that are
clearly not field galaxies, but the field galaxies cannot be uniquely selected that way.

In order to select objects most likely to be field galaxies, we
developed a ``fuzzy boundary'' selection method. From a training set of
objects with known field galaxies, the fraction of field galaxies in a
bin of a relevant SE output parameter can be determined. Our training set 
consists of catalogs of the simulations with no artificial extinction of five galaxies 
in our sample (NGC 1365, NGC 2541, NGC 3198, NGC 3351 and NGC 7331). 
In these catalogs, the added HDF galaxies were identified by their positions 
\footnote{The training set was identified by their positions. The selection of 
field galaxies was based on their properties, not their position.}. The 
fraction of HDF galaxies in a SE parameter bin can then be used as a probability 
that an unknown object with a value in that bin is a field galaxy. By multiplying 
these fractions of HDF galaxies for every relevant SE parameter ($P_i$) for each 
object, an overall galaxy-likeness score (P) for that object is obtained:

\begin{equation}
P = \left({\prod P_i \over (\prod P_i) +( \prod (1-P_i) ) } \right)
\end{equation}

We used the distribution of the log of these probabilities ($log_{10}(P)$) 
as a sliding scale of the galaxy-like quality of an object. 
The distribution of $log_{10}(P)$  for objects in 21 science
fields is plotted in Figure 4, with the distribution of HDF-N/S
objects scaled for comparison. The advantage of using an overall
scale is that an object can fare poorly for one SE parameter but
still make the selection. This makes the boundaries for any single 
parameter in parameter space of the field galaxies ``fuzzy''
\footnote{This resembles a Bayesian approach to the classification 
problem, first applied to star/galaxy separation by \cite{Sebok79}. 
The parameters we use are however not completely independent 
of each other and the HDF percentages are an underestimate of 
the chances as the real galaxies in the bins are not considered 
field galaxies but other objects, skewing the ratio slightly.
 This scoring system however worked well in practice 
for the selection of field galaxy candidates. }. All the structural 
parameters marked in Tables 2 and 3 as well as the V-I color 
from the smallest aperture were used in computing the galaxy score.
The selection criterion is an overall score ($log_{10}(P)$) greater than the $log_{10}$ of the mean score of all objects plus 2.5, that is:

\begin{equation}
log_{10}(P) > log_{10}(P_{mean}) + 2.5 
\end{equation}

Field galaxies missed by this procedure are not selected in either 
simulation or real data and therefore do not influence our comparison.
There are however, still some contaminant foreground objects that 
are selected as well and these have to be identified and discarded by 
visual inspection.

\subsection{Visual identification of contaminants}

A human observer can pick out contaminants based on contextual
information not contained in the SE parameters. There are five broad
categories of remaining contaminants: star-clusters, diffraction spikes, HII
regions, artifacts from dust lanes, and blended objects. 

Stellar clusters, both young open clusters and globular clusters, are
associated with the foreground galaxy. At the distance of the Virgo
cluster, these are often of approximately the size and color of more distant E0
field galaxies. Young open clusters are often found in spiral arms and
are very blue, while globular clusters can be identified by the slightly
different brightness profile. 
Bright stars in our own galaxy result in false selections. Their wings
are extended, often blending with other objects, and the diffraction
spikes resemble edge-on galaxies. The proximity of these false selections
 to the bright star makes them easily visually identifiable.
HII regions resemble blue irregular galaxies but are invariably found in the
proximity of several blue open clusters. 
Dust lanes superimposed on a smooth disk may result in an extended
`object', which is often reddened. This results in severe contamination, 
especially in flocculant spiral galaxies, making their inner regions
unsuitable for the SFM.

Blended objects are by far the largest source of contamination. 
A blend of one of the above objects with a small clump of stars 
is likely to be selected as a candidate field galaxy. Also in a 
nearby foreground galaxy, the granularity of the partly resolved 
disk may result in contamination from blended clumps of disk stars. 
SE does a deblending of peaks in the flux, but the choice of parameters
governing this is a trade-off between deblending objects and keeping
extended objects intact. The candidate objects from the science fields 
were marked in the F814W image for visual inspection together with 
their score and color. Objects deemed to be contaminants were removed. 
All the candidates from the science fields are removed from the synthetic 
field candidate list.

However, the numbers of simulated galaxies have to be corrected for any 
false selections as a result of a blend of a faint HDF object and a foreground 
one. To correct the numbers from the simulated fields, the same visual check 
was done on the simulations from both galaxies with no artificial extinction 
(A=0) and in the case of NGC1365 also in an extincted simulation (A=2). 
The candidate objects were in this case the real galaxies, the simulated 
galaxies and the misidentifications, both from the original field and as a 
result from the addition of the HDF objects. The percentage of HDF 
objects rejected, mostly blends, in these visual checks are given in Table 5 
per typical region, an indicator of the measure of crowding. 
These percentages do not seem to change much as a function of either 
choice of galaxy or simulation. To correct for blends of HDF and foreground objects, 
a fixed percentage of the remaining simulated galaxies from a typical region 
is removed after the removal of the science field's candidates.
These adopted percentages are also given in Table 5 for each typical region.


\section{Improvements in the ``Synthetic Field Method''}

In the process of automating the SFM we have introduced several improvements. 
First, exposures were combined with the ``drizzle'' routine, improving the 
sampling of the final image. Second, we have provided for a less observer 
dependent selection of field galaxies. These two categories of improvements 
we have described in the previous sections. Third, extra simulations were 
made, biases and uncertainties were estimated, and opacities were obtained 
based on segments of the images with similar characteristics. We will describe 
these improvements in this section.

\subsection{Foreground galaxy segmention}

The SFM provides an average opacity for a certain 
region of the foreground galaxy. \cite{Gonzalez98} reported opacities 
for regions defined by WFPC2 chip boundaries. Ideally, an average 
opacity is determined for a region of the foreground galaxy that is 
homogeneous in certain characteristics: arm or inter-arm regions, 
deprojected radius from the center of the galaxy, or a region with the 
same surface brightness in a typical band.

In our treatment, the mosaiced WFPC2 fields are visually devided into 
crowded, arm, inter-arm and outside regions. This step is applied to the 
catalogs of objects by tagging each object according to its general location 
in the foreground galaxy. Objects from the crowded regions were ignored 
in the further analysis. 
The deprojected radial distance for each object was also computed from 
the inclination, position angle and position of the galaxy center taken from 
the 2MASS Large Galaxy Atlas \citep{LGA} or, alternatively, the extended 
source catalog \citep{extended}; the distance was taken from \cite{KeyProject}. 
The surface brightness based on the HST/WFPC2 mosaics or a 2MASS 
image could also be used to define a partition of the WFPC2 mosaics.

\subsection{Simulated fields}

Simulated fields are made by taking one WF chip from either the northern 
or southern Hubble Deep Field (HDF-N or HFD-S), extincting it with a 
uniform grey screen, and adding it to a data WF chip. This results in six 
separate simulations for each opacity and data chip: one for each HDF-N/S 
WF chip. Simulations for seven opacity levels were made, ranging from 
-0.5 to 2.5 magnitudes of extinction with steps of 0.5 magnitude. The 
negative -0.5 opacity simulation was added to obtain a more accurate 
fix on the point of zero opacity. The use of a grey screen in the simulations 
was chosen because its effect on the numbers of field galaxies is similar 
to that of a distribution of dark, opaque clouds with a specific filling factor 
and size distribution. \footnote{Moreover, \cite{Gonzalez98} found that 
assuming a Galactic or a grey extinction curve made no difference in the 
extinction derived in NGC 4536 using the SFM.}

To infer the opacity ($A_I$) from the numbers of field galaxies ($N$), \cite{Gonzalez98} use:

\begin{equation}
A_I = -2.5 ~ C ~ log \left( N \over N_0 \right)
\end{equation}

$N_0$ is the normalisation and C the slope of the relation between the number of field galaxies and the extinction. They depend on the crowding in the field and total solid angle. Crowding limits the number of field galaxies. When it dominates the loss of field galaxies, the relation becomes much flatter ($C>>1$). \cite{Gonzalez98} found C to differ with the extinction law used in the simulations in the same foreground field. We use grey extinction but vary the foreground field.

For each field, we fit the relation between $A_I$ and $log(N)$, minimising $\chi^2$  to the average numbers of field galaxies found in the simulated fields with known extinctions \footnote{N is the average of HDF-N and HDF-S as a reasonable approximation of the number of galaxies expected from the average field. The possible deviation of actual background field from the average is accounted for in the error estimate.}.
The intersection of this curve with the real number of field galaxies yields an average opacity estimate for the region. See Figures 6, 9, 10 and 11 for the fitted relation (dashed line) and the number of field galaxies from the science field (solid line).

\subsection{Field galaxy numbers: uncertainties and systematics}

There are four quantities which affect the numbers of field galaxies besides 
dust absorption. They are: crowding, confusion, counting error and clustering. 
Crowding and confusion introduce biases which need to be calibrated. Counting and clustering introduce uncertainties in the galaxy numbers that must be estimated. In addition, the clustering could possibly introduce a bias if the reference field is not representative for the average.\\
\noindent{\it Crowding} effectively renders the parts of the image of little use for the SFM. 
Typically these are stellar clumps, the middle of spiral arms and the center of the foreground galaxy.
The strongly crowded regions in the WFPC2 mosaics were masked off
and not used in further analysis.\\
{\it Confusion} is the misidentification of objects by either the selection algorithm or the observer. Misidentification by the algorithm is corrected for by the visual check of the science fields (detailed in section 3.4). In order to correct the numbers of simulated objects, the candidates from the science field, including the misidentifications, are removed and subsequently the average rejection rate from Table 5 is applied to the remaining objects from each typical region. The typical regions are a measure for the crowding, the main source of the remaining confusion due to blends of HDF and foreground objects. \\
\noindent {\it Counting} introduces a Poisson error. If the numbers are 
small ($N<100$), $\sqrt{N}$ underestimates the error and the expressions 
by \cite{Gehrels85}  for upper and lower limits are more accurate. 
We adopted these for both simulated and real galaxy numbers using the expressions for upper and lower limits for 1 standard deviation.\\
{\it Clustering} introduces an additional uncertainty in the number of {\it real} 
galaxies in the science fields, as the background field of galaxies behind the foreground galaxy 
is only statistically known. This variance in the background field necessitates a prudent choice of reference field for the background in the simulated fields as otherwise an inadvertent bias in the opacity measurement can be introduced (see also section 4.3.1).

The standard deviation of this uncertainty can be estimated using a similar argument 
to the one in \cite{Peebles80} (p.152), replacing volume by solid angle and the three 
dimensional 2-point correlation function by the two dimensional one ($\omega(\theta)$). 
The resulting clustering uncertainty depends on the depth of the observation and 
the solid angle under consideration.

\begin{equation}
\sigma^2_{clustering} = N + N^2 \times \left(A(m_{lim},Filt) 
{2 \Gamma (2+\delta \over \Gamma (2+ {\delta \over 2})  \Gamma (3+ {\delta \over 2}) }
\theta_{max}^{\delta} \right)
\end{equation}

where $A(m_{lim},Filt)$ is the amplitude, depending on photometric band and brightness interval and  $\delta$ is the slope of the 2-point correlation function
$\omega(\theta) = A ~ \theta^{\delta}$. N is the number of field galaxies and $\theta_{max}$ characterises the size of the solid angle under consideration. 
The slope, $\delta$ is usally taken to be -0.8 and the value of term between $A(m_{lim},Filt)$ and $\theta_{max}^{\delta} $ in equation 4 becomes 1.44.
The $A(m_{lim},Filt)$ values from \cite{Cabanac00} are used to compute $\omega(\theta)$ 
and the resulting clustering uncertainty as they are for the same filters (V and I) and integrated over practical brightness intervals with a series of limiting depths
\footnote{Although two point correlation functions have been published based on HST data, these  results are for very narrow magnitude ranges (see the references in Figure 5.) 
The results from \cite{Cabanac00} are for similar magnitude ranges as the objects from our crowded fields and are given for the {\it integrated} magnitude range.}.
\cite{Gonzalez98} claim a completeness of galaxy counts up to 24 mag for one of their fields. 
We estimate the limiting magnitude by the value above which 90\% of the simulated galaxies with no dimming (A=0) lie.
We extrapolated the relation between limiting magnitude and amplitude ($A(m_{lim},Filt)$) from \cite{Cabanac00} to model the clustering error to higher limiting magnitudes. For each field we characterise the limiting depth by the interval in which the majority of simulated field galaxies lie in the simulations with no opacity.
Alternatively we could have used a very large number of background fields 
in the simulations and determined the possible spread in field galaxy numbers due to clustering 
from those. For practical reasons we used the average of simulations with the HDF-N/S fields and estimated the uncertainty in the real number of field galaxies from equation 4. The uncertainties in opacity owing to the clustering uncertainty in the original background field are given separately in Table 6, and in Figures 6, 9, 10 and 11.

The clustering error and Gehrels's counting uncertainties were added in quadrature 
to arrive at the upper and lower limits of uncertainty for the real galaxies. 
Simulated counts only have a counting uncertainty as these are from a known 
typical background field.

From the errors in the number of field galaxies in each simulation and in the real 
number of field galaxies, the uncertainty in the average opacity A can then be derived 
from equation 3.
A single field gives a highly uncertain average value for extinction. Averaging 
over several galaxies will improve statistics and mitigate the error from 
field galaxy clustering.

\subsubsection{The HDF as a reference field}

The SFM uses the Hubble Deep Fields as backgrounds in the synthetic fields. The counts from these are taken to be indicative of the average counts expected from a random piece of sky, suffering from the same crowding issues as the original field. In this use of the HDF-N/S as the reference field, the implicit assumption is that it is representative of the average of the sky. If they are not, the difference in source counts between the HDFs and the average sky introduces a bias in the synthetic fields and hence a bias in the resulting opacity measure. 

The position of the HDF North was selected to be unremarkable in source counts and away from known nearby clusters \citep{HDFNobs}. The position of the HDF South was dictated by the need to center the STIS spectrograph on a QSO but \cite{HDFSobs} assert that the source count in HDF-S was unlikely to be affected by that. In addition \cite{HDFS} point out that the HDF-S was chosen such that it was similar in characteristics to HDF-N. The selection strategy of the Deep Fields therefore does not seem to be slanted towards an overdensity of sources.

To test the degree to which the Hubble Deep Fields are representations of the average field of sky, the numbers of galaxies we find can be compared to numbers from the Medium Deep Survey \citep{MDS}, a program of parallel 
observations with the WFPC2, also in F814W.  Several authors \citep{Casertano95,Driver95a,Driver95b,Glazebrook95,Abraham96,Roche97} report numbers of galaxies as a function of brightness in these fields. \cite{Casertano95, Driver95a, Glazebrook95, Abraham96, Roche97} present averages for multiple fields and \cite{Driver95b, Abraham96} the numbers from deep fields, the latter for the HDF-N. In Figure 5, we plot these numbers of galaxies as a function of magnitude. Also the number of sources identified by our algorithm as field galaxies in the Hubble Deep Fields are also plotted. The average of the Hubble Deep Fields (filled circles) corresponds well to the curves from the literature up to our practical limiting depth of 24 mag. The difference between the north and south HDF never exceeds the Poisson uncertainty of the average, and even changes sign 
for galaxies fainter then 24 mag.
 
From Figure 5, we conclude that the average of the Hubble Deep Fields is a good representation of the average field in the sky, and the numbers of galaxies from the simulations do not need to be corrected for  any bias resulting from an atypical reference field. In any case, any residual bias would be trivial compared to the uncertainties in individual WFPC2 fields, but could have become important when combining counts from many fields as we have done in our companion paper \citep{Holwerda04b}.

\subsection{Inclination correction}

Any inclination correction of the opacity values depends on the assumed dust geometry. A uniform 
dust screen in the disk would result in a factor of cos(i) to be applied to the opacity 
$A_I$. However if the loss of field galaxies is due to a patchy distribution of opaque 
dust clouds, the correction becomes dependent on the filling factor, cloud size 
distribution and cloud oblateness. 
All the extinction estimates ($A_I$) and the values corrected for inclination ($A_I \ \times \ cos(i)$) are listed in Table 6, assuming a simple uniform dust screen.

\section{Examples: NGC 4536 and NGC 1365}

NGC 4536 \citep{Gonzalez98} was reanalyzed as a test case to 
provide a comparison between observers and versions of the 
SFM. NGC 1365 was one of the first galaxies analyzed with the improved method \citep{Holwerda02} 
and provides a good example of how the method works for an image which can be segmented into different regions. See Table 4 for basic data on both galaxies and the observations which made up the data set.

\subsection{NGC 4536, comparing observers}

Identifying field galaxies remains a subjective process, and different software systems as well as different observers will differ in their identifications. However, as long as the same selection criteria are applied to simulations and science objects , a good estimate of dust extinction can be made. Figure 6 shows the extinction measurements from \cite{Gonzalez98} and this paper.

The numbers of field galaxies and subsequent derived opacities in the combined WF2 and 3 chips are very similar for \cite{Gonzalez98} and this paper. The results for WF4 seem to differ however, both in numbers of field galaxies found as the derived extinction. WF4 was analysed separately by \cite{Gonzalez98} as it was less crowded than the other two WFPC2 chips, so field galaxies could be found to a higher limiting depth. \cite{Gonzalez98} estimate the limiting magnitude for the WF4 chip as 24 magnitude and for the WF2\&3 field as 23 mag. The selection of objects as candidate field galaxies by our algorithm however imposes a limiting depth to which objects are selected ( $I \approx 23 mag$). This effect can be seen in the cumulative histogram of real galaxies (Figure 7), especially in the WF4 where the numbers from \cite{Gonzalez98} are still increasing beyond 24 magnitude. The effect is less pronounced in the simulated numbers (Figure 7 bottom right). These differences in numbers at the faint end are the cause for the difference in derived opacities for \cite{Gonzalez98} and this paper. However a lower limiting magnitude makes the derived extinction less accurate (lower number statistics and a bigger uncertainty due to clustering), not inconsistent with each other. 

If the numbers from \cite{Gonzalez98} are limited to the same limiting magnitude as ours, the numbers from science and simulated fields galaxies match up. In addition, in a visual check, both observers agree on the identification of these brighter field galaxies. Given this, we feel that the automated method's trade of depth for speed is warranted. 

By automatically selecting objects and correcting the simulations for the pruning of galaxies in the visual step, we are confident that we select similar sets of field galaxies, to the same limiting depth, with a high degree of certainty in both simulated and real fields.

\subsection{NGC 1365: arm and inter-arm extinction}

\cite{Gonzalez98} remarked on the importance of distinguishing between 
arm regions, regions between the arms (inter-arm), and outside regions. 
 \cite{Beckman96}, \cite{kw00a} and \cite{kw00b} all found that 
extinction was more concentrated in the spiral arms. 
NGC 1365 provides a nice example of an arm with a crowded region, an 
inter-arm region, a spur, and some outside area (see Figure 8). The opacity 
measurements of these regions are plotted in Figure 9. The ``spur'' region 
(IV), the inter-arm region (III) and the region outside (V) show some opacity 
but all are still consistent with none. Most of the extinction in this galaxy is 
in the main inner arm (II): $A = 3.9_{-3.5}^{+2.6}$. 
For such a small subdivision in only one field, this opacity measure is 
very uncertain.  However, by combining measurement in several arm 
regions in several galaxies, as we have done in \cite{Holwerda04b}, we are 
confident that a reliable and meaningful estimate can eventually be made.

\subsection{NGC 4536 and NGC 1365: Radial profile of extinction}

One of the new applications of the SFM introduced here is to compare the 
numbers of field galaxies in an annular region of the mosaic between two 
deprojected radii. Figures 10 and 11 show results for four sets of annuli for 
NGC 4536 and NGC 1365, respectively. We present the radial opacity 
values found in this way in Figure 12 for both NGC 4536 and NGC 1365. 
Noticeable is the occurrence of comparatively high values of opacity at 
different radii. This depends on whether or not the area in the radial annulus 
is dominated by arm regions or inter-arm type regions. 

The NGC1365 profile shows a steep rise in the inner region and the peak 
in the NGC 4536 profile corresponds to the prominent 
arm there. Individual errors in these measurements remain quite large due 
to the poor statistics and the clustering uncertainty in the field of galaxies.

Comparing these values to those in Figure 12 of \cite{kw00a}, the peak 
values in these radial plots, at 0.3 and 0.75$R_{25}$ respectively (see Figure 12), 
are completely consistent with the arm extinction values found by those authors.
When combining the radial profiles of our entire sample of HST fields, 
we should keep in mind the importance of spiral arms in the radial extinction profile. 

\subsection{Surface Brightness}

\cite{Giovanelli95, Tully98, Masters03} found that disk extinction correlates 
with total galaxy luminosity. With the SFM we can have a more detailed look 
at the correlation between the light in a galaxy and the extinction. The higher 
extinction found in spiral arms is an indication that this correlation is also 
present in our data. However, with few points obtained from only two fields, 
no relation can be reliably detected. A plot of the extinctons and surface 
brightnesses derived from all our fields will be presented in a future paper.

\section{Conclusions}

We have shown that the Synthetic Field Method as developed by
\cite{Gonzalez98} can be successfully automated and applied to a
large variety of fields. As most classification schemes break down in
crowded regions, some visual check by a human observer will remain
necessary, either to deem the region too crowded or to check the classification of the
objects. The bias thus introduced is also calibrated using synthetic fields.
The great increase in throughput provided by the automation opens up 
the possibility to infer dust absorption in a wide range of fields available 
in the HST archive.

In the process of automating the SFM, we introduced some
improvements. The quality of the images has been improved with
the drizzle technique. The selection of field galaxies is
less observer-dependent and much faster. Extra HDF-S control fields were
added to mimic the average background field. The results are now given 
per typical region instead of per chip. Improved estimates of the uncertainties 
due to the random error and the clustering of the field galaxies have also
been incorporated.

Future improvements of this technique could include the use of 
multi-color imaging or field spectroscopy in order to more unambiguously 
identify the field galaxies. An improved object classification, based 
on different data and with a more sophisticated algorithm, could in 
the future make the need for a visual check of objects redundant.

The apparent difference between derived extinctions between this paper 
and \cite{Gonzalez98} for NGC4536 can be accounted for by a  
difference in limiting depth and the uncertainty it brings with it. 
However the agreement between observers in their identifications of 
the brighter objects suggest a consistency of the method across 
identification schemes. 

The radial dependencies of opacity in our examples show evidence 
of substantial extinction, $A_I = 0.6_{-0.7}^{+0.6}$ mag for NGC 1365 
at half the $R_{25}$ and $A_I = 1.6_{-1.3}^{+1.0}$ at 0.75 $R_{25}$ 
radii for NGC 4536. These extinction values at these radii are consistent 
with those reported by \cite{kw00a}. Most of this extinction 
seems to concentrate in the arm regions of these galaxies.

While the SFM itself is independent of assumptions about the dust geometry 
in the foreground galaxy, the inclination correction for the opacity is not. 
Corrections based on a simple screen have been presented, a more thorough 
discussion of other possibilities is considered in the companion paper \citep{Holwerda04b}.

The principal advantage of this method is that no assumption about the 
distribution of either absorbers or the underlying starlight goes into the 
measurement. However, the small number statistics in individual regions 
result in large uncertainties for single measurements. Averages over 
several fields will improve this by increasing the statistics and averaging 
out the field galaxy clustering. The application of this method to a substantial set
 of archival HST/WFPC2 images is presented in \cite{Holwerda04b}

\section{Acknowledgements}

We would like to thank A. Koekemoer for his help with
the ``drizzle'' method, E. de Vries and I. Smail for help with modifying the
Source Extractor code, and H. Ferguson and S. Casertano for insightful discussions
on the Synthetic Field Method. Earlier versions of this manuscript were read by
E. Brogt.

This research has made use of the NASA/IPAC Extragalactic Database 
(NED) which is operated by the JPL, CalTech, under contract with NASA.

This work is primarily based on observations made with the NASA/ESA Hubble Space Telescope and
obtained from the data archive at the Space Telescope Institute. 
Support for this work was provided by NASA through grant number HST-AR-08360 from the Space Telescope Science Institute (STScI).
STScI is operated by the association of Universities for Research in Astronomy, 
Inc. under the NASA contract  NAS 5-26555.
We are also grateful for the financial support of the STScI Director's Discretionary Fund (grant numbers 82206 and 82304) and the Kapteyn Institute of Groningen University

\begin{figure}
\plotone{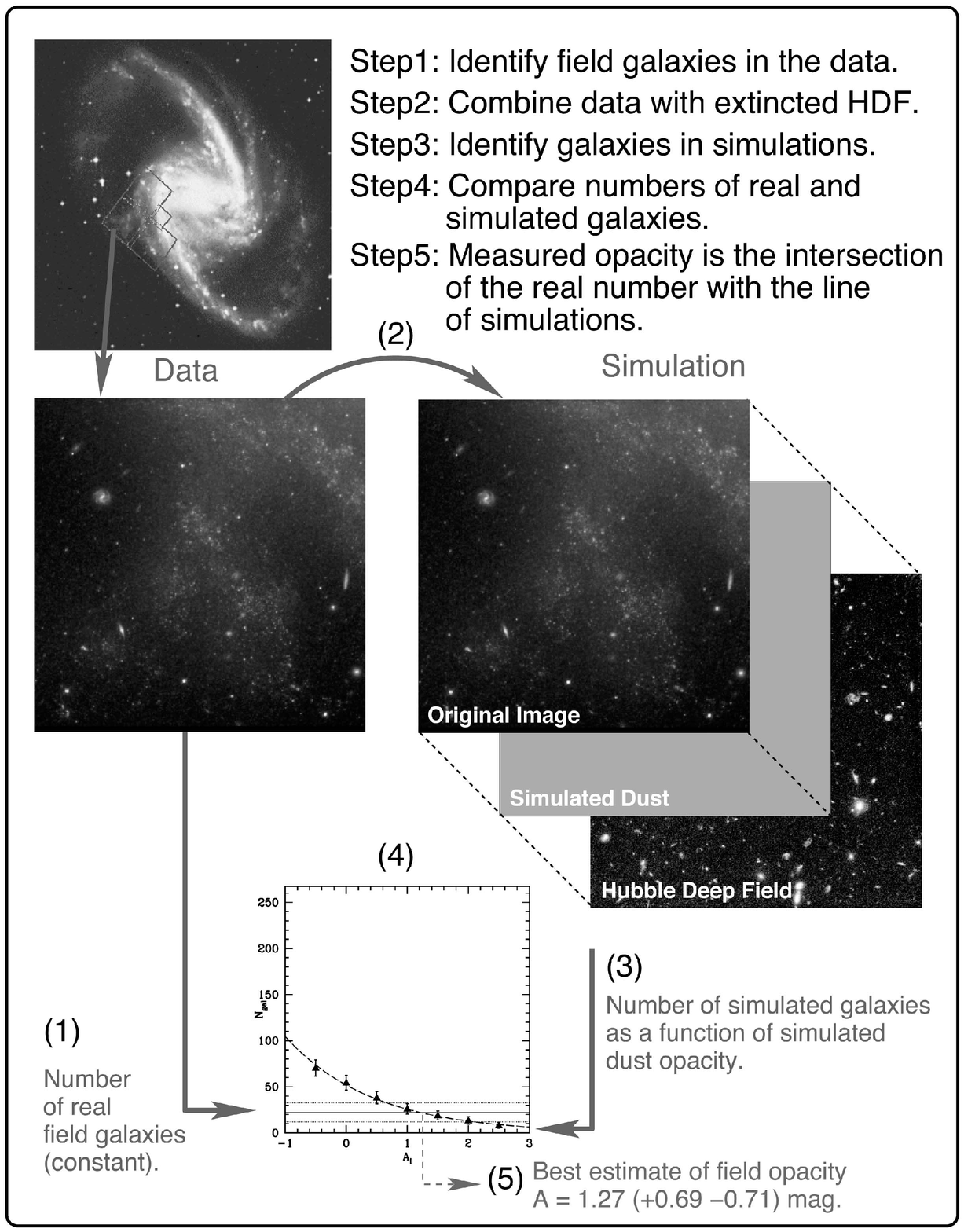}
\caption{A schematic of the Synthetic Field Method (SFM). First, field galaxies are identified in the science field by a combination of automatic and visual selection. Secondly, a a HDF field is added to the science field in a series of simulations with different opacities. Field galaxies are selected from these simulated fields. Eq. 3 is fitted to these and uncertainties are estimated.
Finally, the intersection between that relation and the number of galaxies gives the opacity of the area under consideration. In this case, the WF3 chip of NGC1365 has an average extinction of $1.3^{0.7}_{-0.7}$ magnitude.}
\end{figure}

\begin{figure}
\plotone{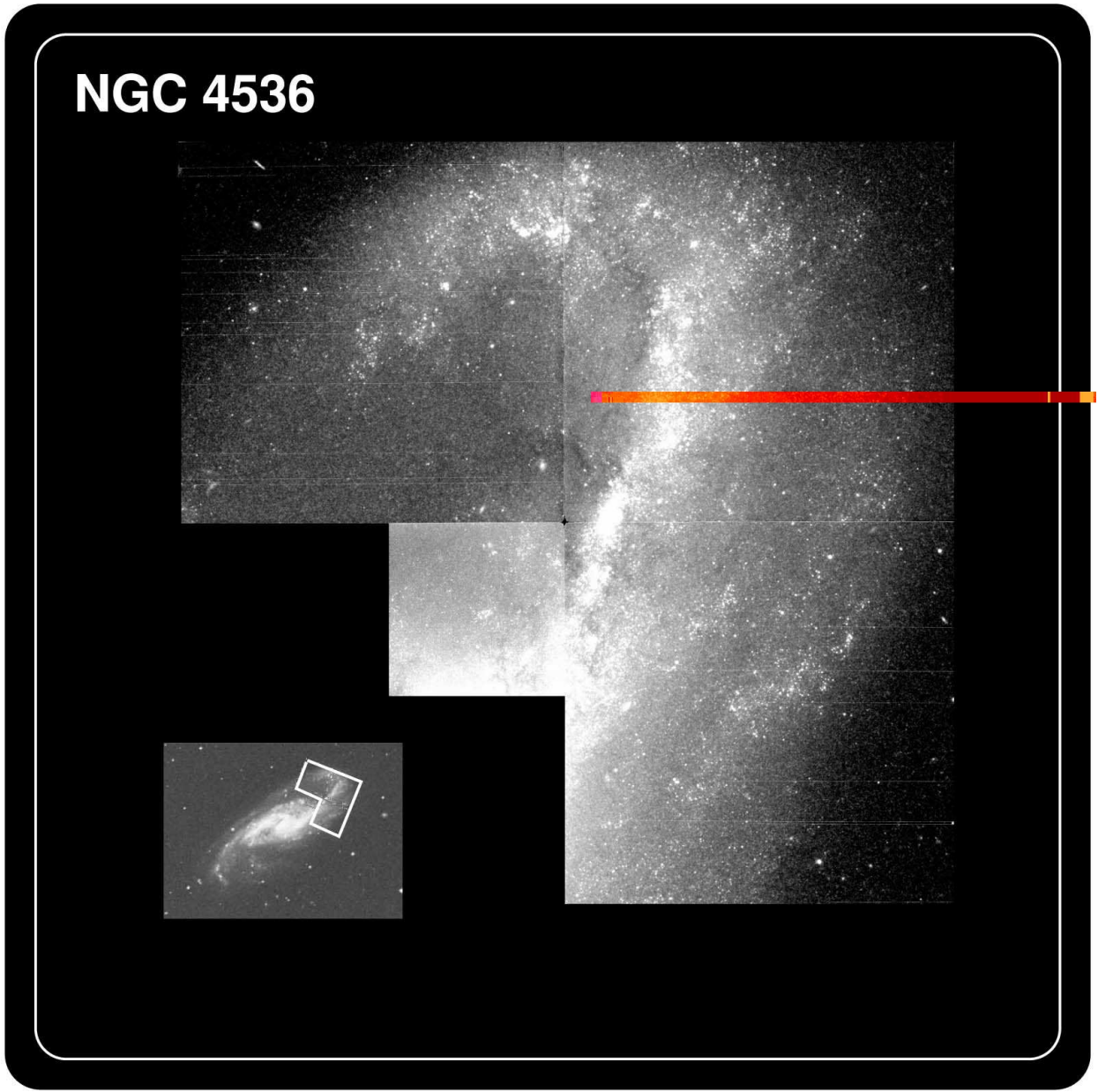}
\caption{The mosaic made of the HST/WFPC2 exposure of the NE arm of NGC 4536, the inset is the Digital Sky Survey image with the WFPC2 footprint. Because of the lack of a dither greater then one pixel between epochs, some of the effects of masked bad columns can still be seen.}
\end{figure}

\begin{figure}
\plotone{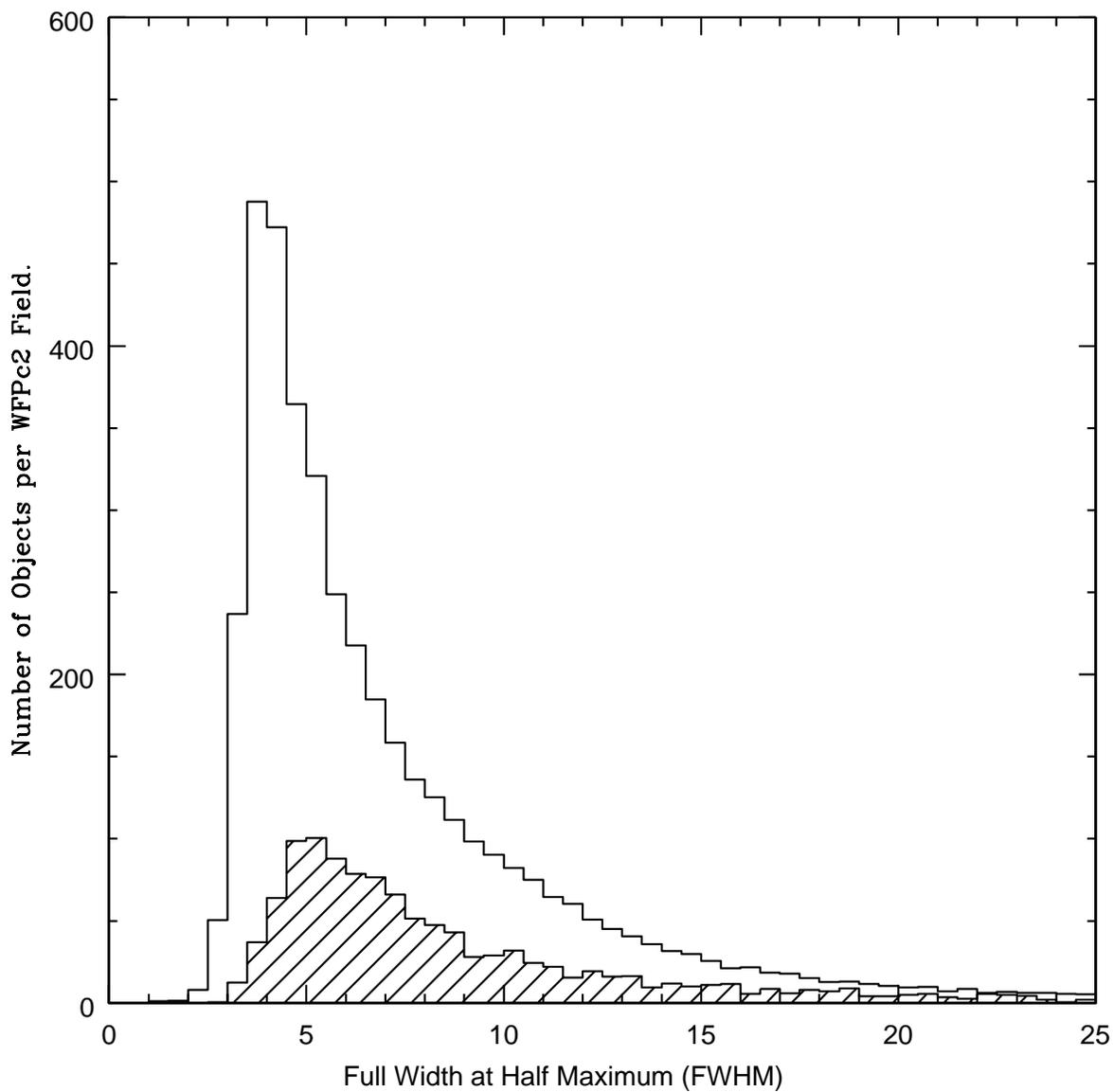}
\caption{The distribution of the FWHM (in pixels), determined by Source Extractor, of all the
objects in 21 of the science fields, averaged over the number of
fields. The shaded area is the histogram per WFPC2 field for HDF
galaxies (both North and South). A selection limit based on this
parameter only would not have done nearly as well as our scoring system.}
\end{figure}

\begin{figure}
\plotone{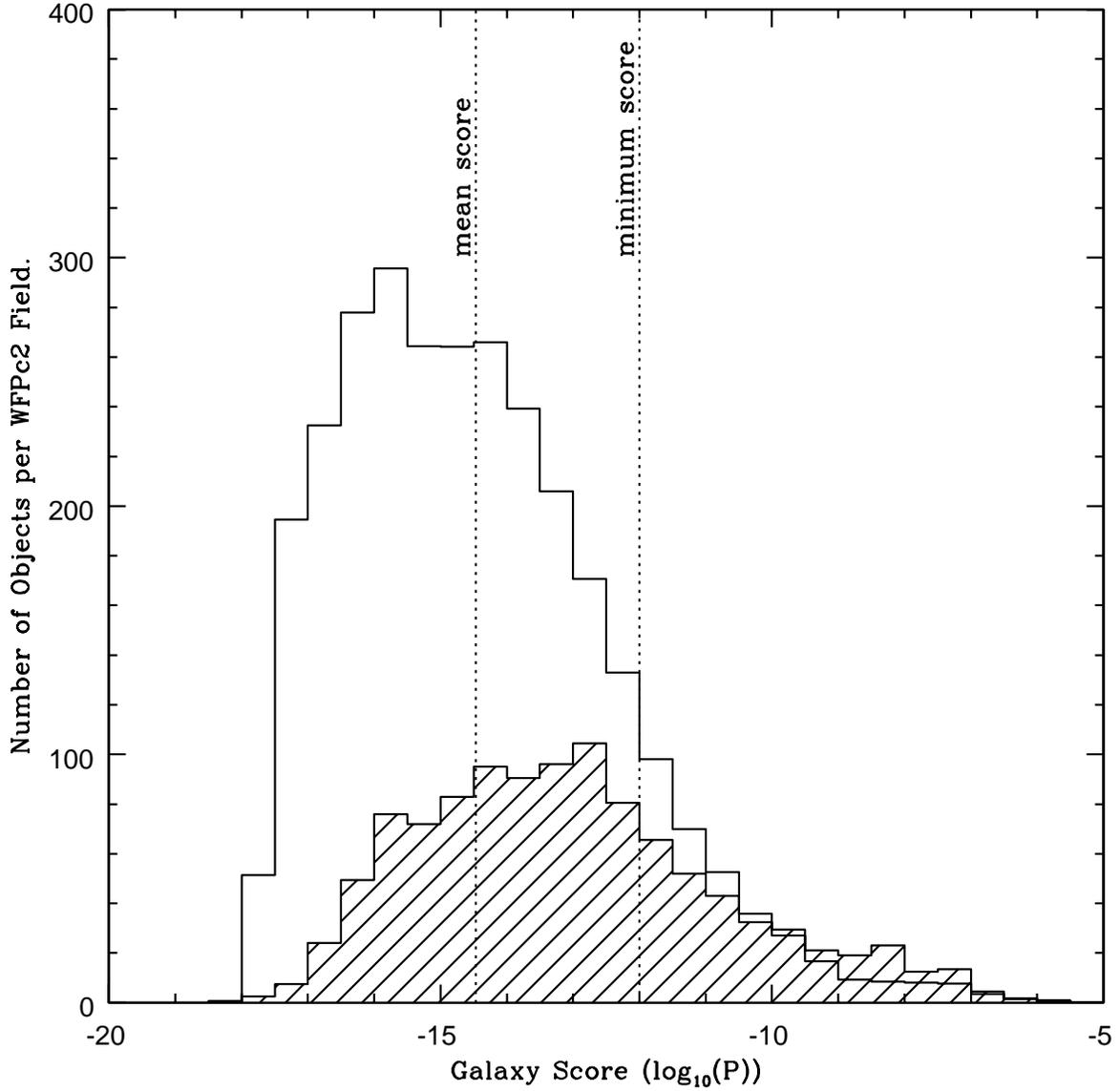}
\caption{
The distribution of galaxy score for objects in
our science fields. $P = \left({\prod P_i \over (\prod P_i) +( \prod (1-P_i) ) } \right)$. 
The shaded area is the average histogram for HDF
galaxies (both North and South). Objects to the right are more galaxy-like. The majority of field galaxies is indistinguishable in properties from the objects in the foreground galaxy. Only the higher scoring tail (about -12 and above) can be used for the opacity
measurement. The mean score of all objects is indicated, together with the minimum score for selection.}
\end{figure}

\begin{figure}
\plotone{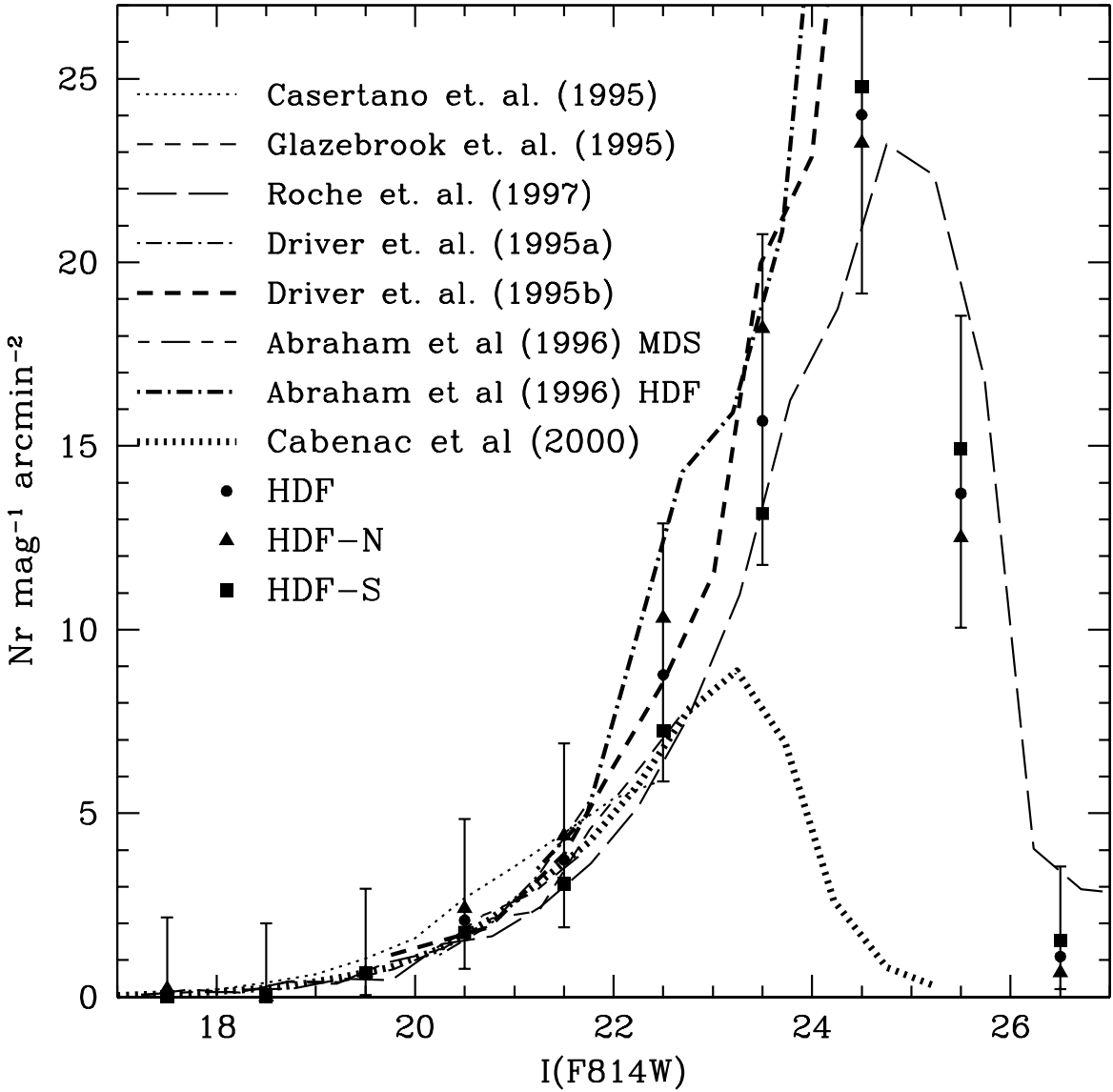}
\caption{The numbers of field galaxies per magnitude per square arcminute from several authors. All counts were in the I band (F814W). The HDF points are the average of all our counts for HDF-N and HDS-S. Triangles and squares are our averages for HDF-N and HDF-S respectively. The numbers by \cite{Casertano95} are from pre-refurbished WFPC data, accounting for the slightly higher numbers. The numbers found by \cite{Abraham96} for the HDF-N are slightly higher then ours for the fainter objects as our selection started to discard some. The dashed lines are from single deep WFPC2 exposures \citep{Driver95b,Abraham96}, the solid lines from multiple exposures \citep{Casertano95,Glazebrook95, Driver95a, Abraham96, Roche97}.   }
\end{figure}

\begin{figure}
\plotone{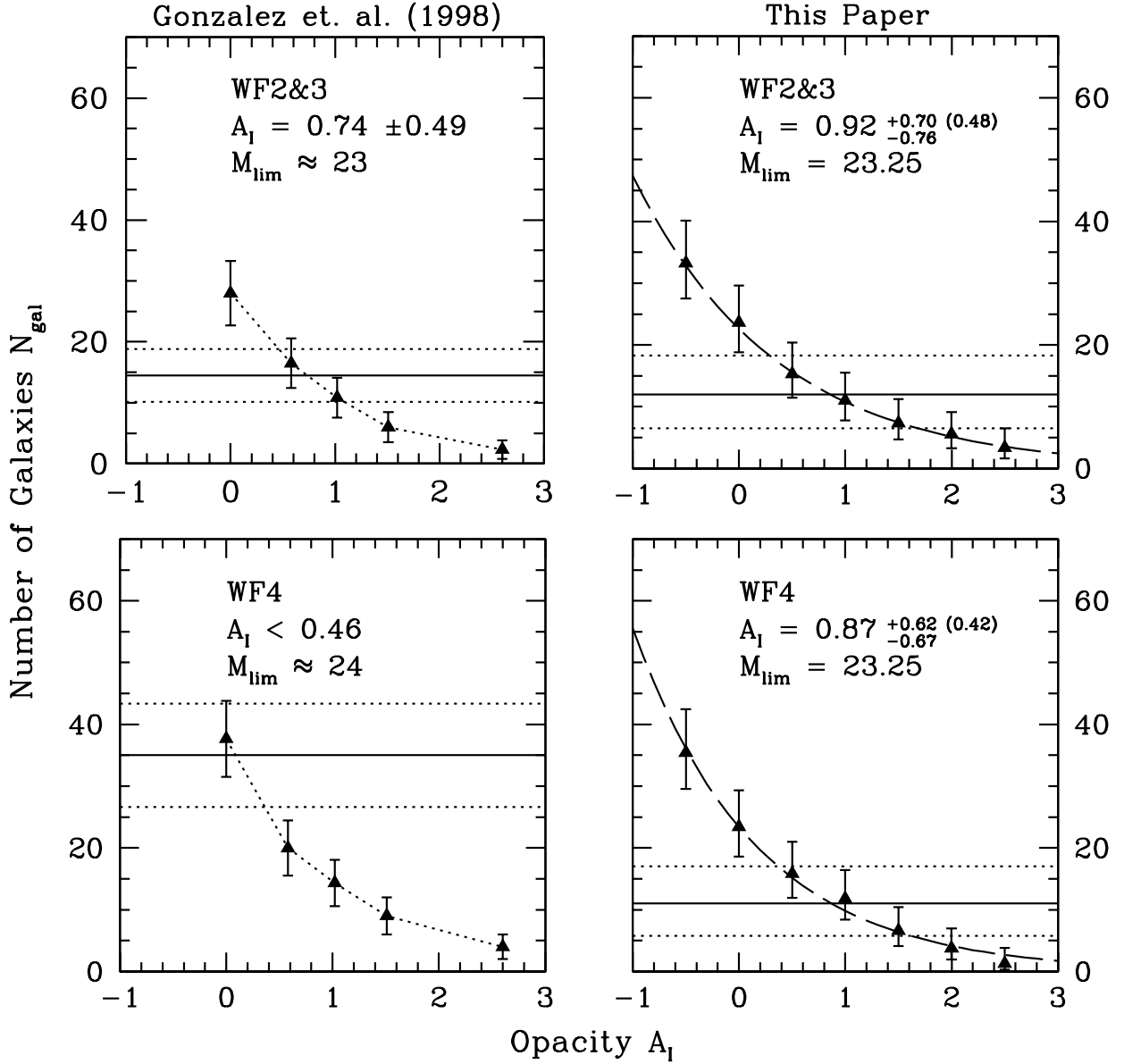}
\caption{The number of simulated galaxies per WF chip as a function of extinction for the \cite{Gonzalez98} result (left panels) and this paper (right panels). Top panels are the average for WF2 and 3, combined by \cite{Gonzalez98} due to similar appearance and poor statistics. The bottom panels are WF4. The errorbars for the simulated numbers (filled triangles) are Poisson uncertainties only. The dashed line is the best fit ($A_I = -2.5 \ C \ log_{10}(N / N_0)$). 
The solid horizontal line is the real number of field galaxies found, the dotted horizontal lines mark the uncertainty in this number due to counting and clustering combined. The opacity measurement shows also the total error, and that part of the error which is due to clustering in brackets. The limiting magnitude $M_{lim}$ was determined from the A=0 simulation. }
\end{figure}

\begin{figure}
\plotone{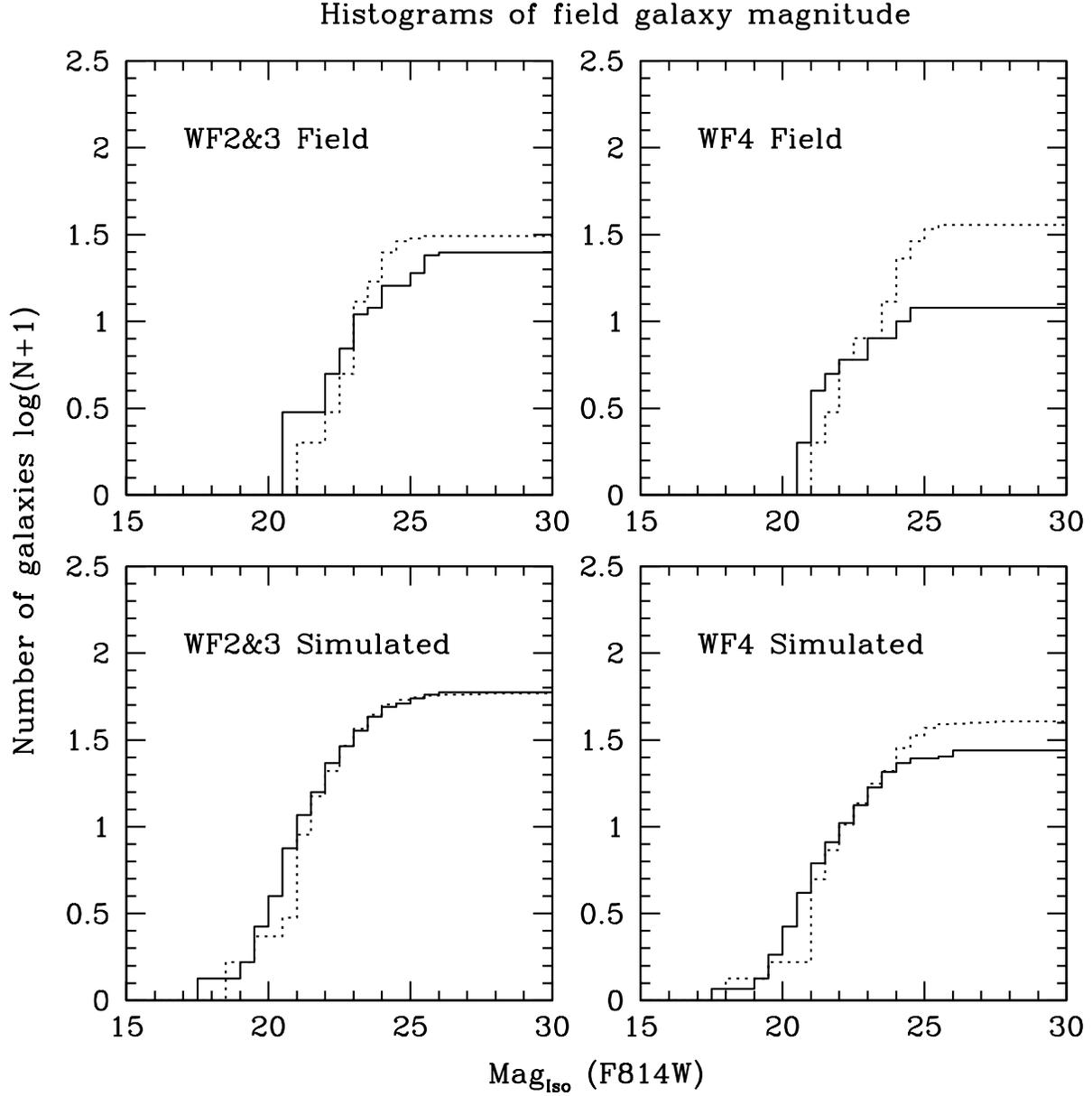}
\caption{The cumulative histograms of number of field galaxies with their magnitude (MAG\_ISO) for our identifications (solid lines) and \cite{Gonzalez98} (dotted lines) for wf2\&3 (left panels) and wf4 (right), science field galaxies (top) and simulated field galaxies (bottom). }
\end{figure}

\begin{figure}
\plotone{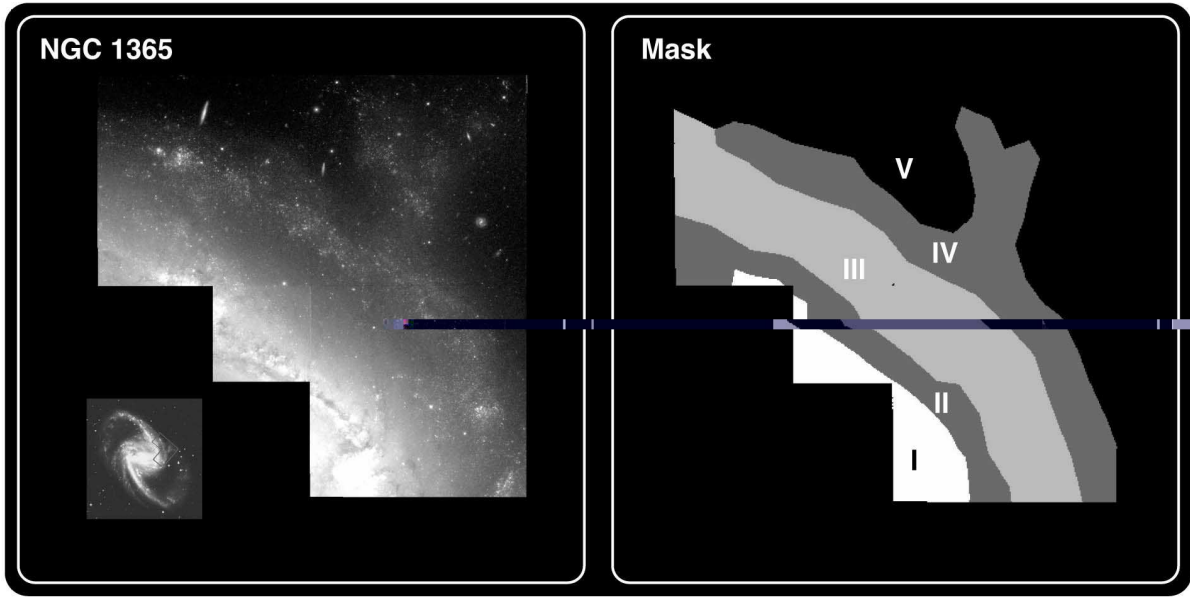}
\caption{The mask used to denote crowded (I), arm (II, IV), inter-arm (III) and outside (V) regions in NGC 1365. Galaxy number counts are given for the inner arm region (II), inter-arm region (III), the ``spur'' (IV), and the outside region (V) in Figure 9. }
\end{figure}

\begin{figure}
\plotone{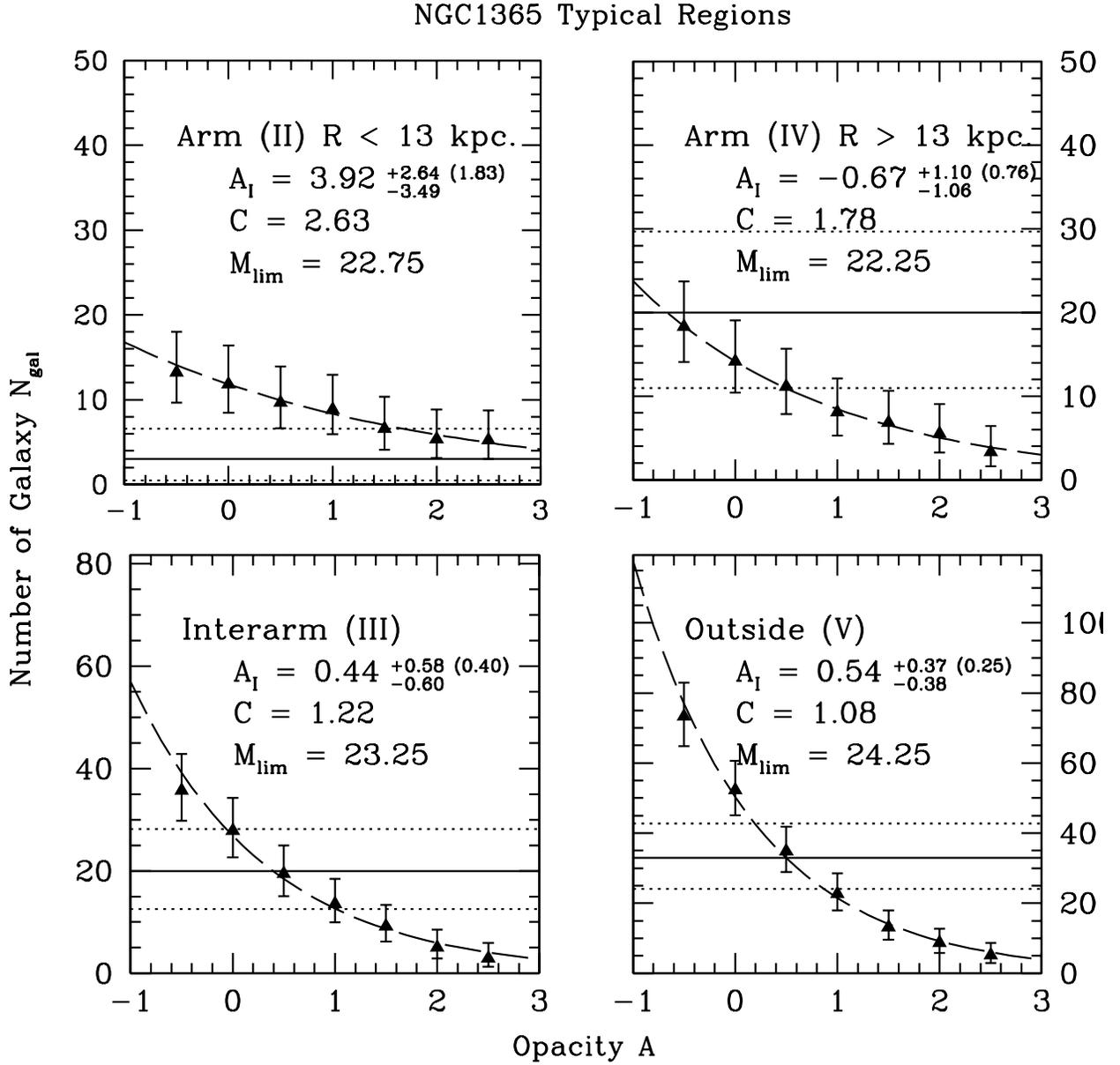}
\caption{Simulated and real numbers of field galaxies in typical regions in NGC 1365;  ``arm'' regions on either side of the inter-arm region (region II and IV in figure 7), the ``inter-arm'' region (region III) and the ``outside'' region (region V). Errorbars, notation and curves same as the right panels in Figure 6. C is the fit for equation 3. $M_{lim}$ is the limiting magnitude used to compute the clustering uncertainty.}
\end{figure}

\begin{figure}
\plotone{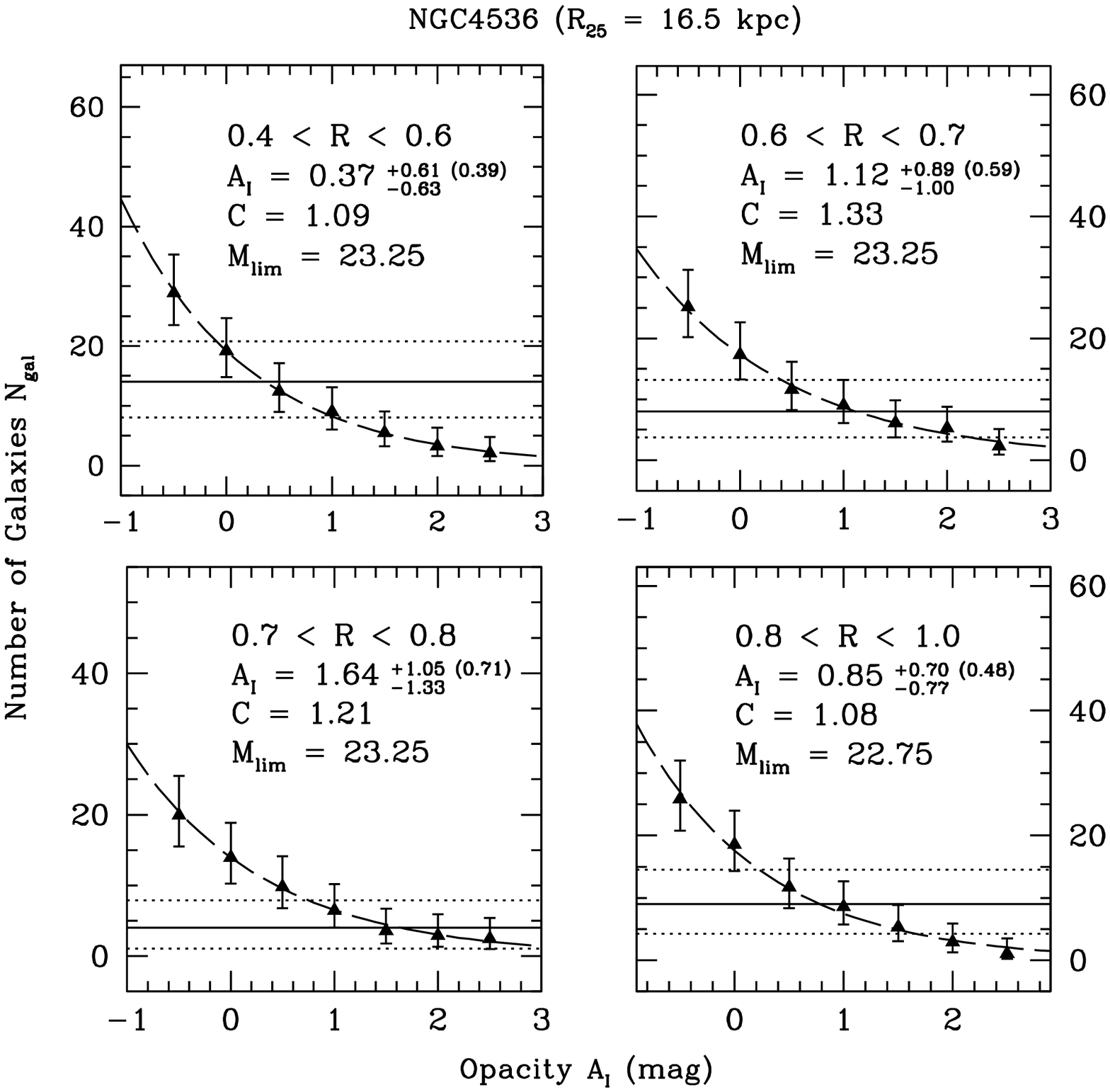}
\caption{The numbers of real field galaxies and those of simulated field galaxies plotted as a function of simulated opacity in areas of de-projected radius from the centre of NGC 4536. The radii are expressed in $R_{25}$, derived from the B-band photometric diameter ($D_{25}$) from the RC3 catalog \citep{RC3}. Errorbars, notation and curves same as the right panels in Figure 6. C is the fit for equation 3. $M_{lim}$ is the limiting magnitude used to compute the clustering uncertainty.}
\end{figure}

\begin{figure}
\plotone{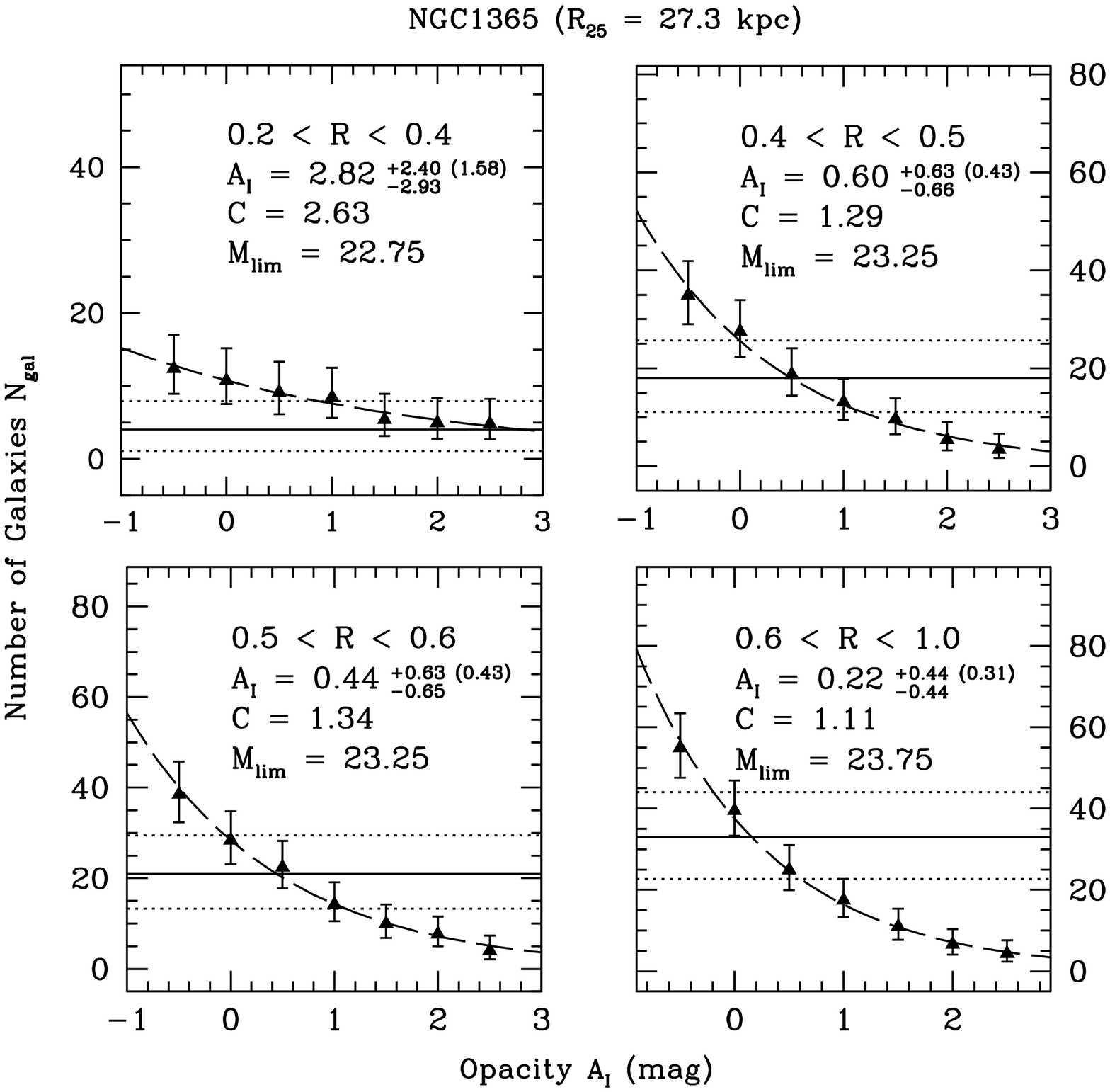}
\caption{The numbers of real field galaxies and those of simulated field galaxies plotted as a function of simulated opacity in areas of deprojected radius from the center of NGC 1365. The radii are expressed in $R_{25}$, derived from the B-band photometric diameter ($D_{25}$) from the RC3 catalog \citep{RC3}. Errorbars, notation and curves same as the right panels in Figure 6. C is the fit for equation 3. $M_{lim}$ is the limiting magnitude used to compute the clustering uncertainty.}
\end{figure}

\begin{figure}
\plotone{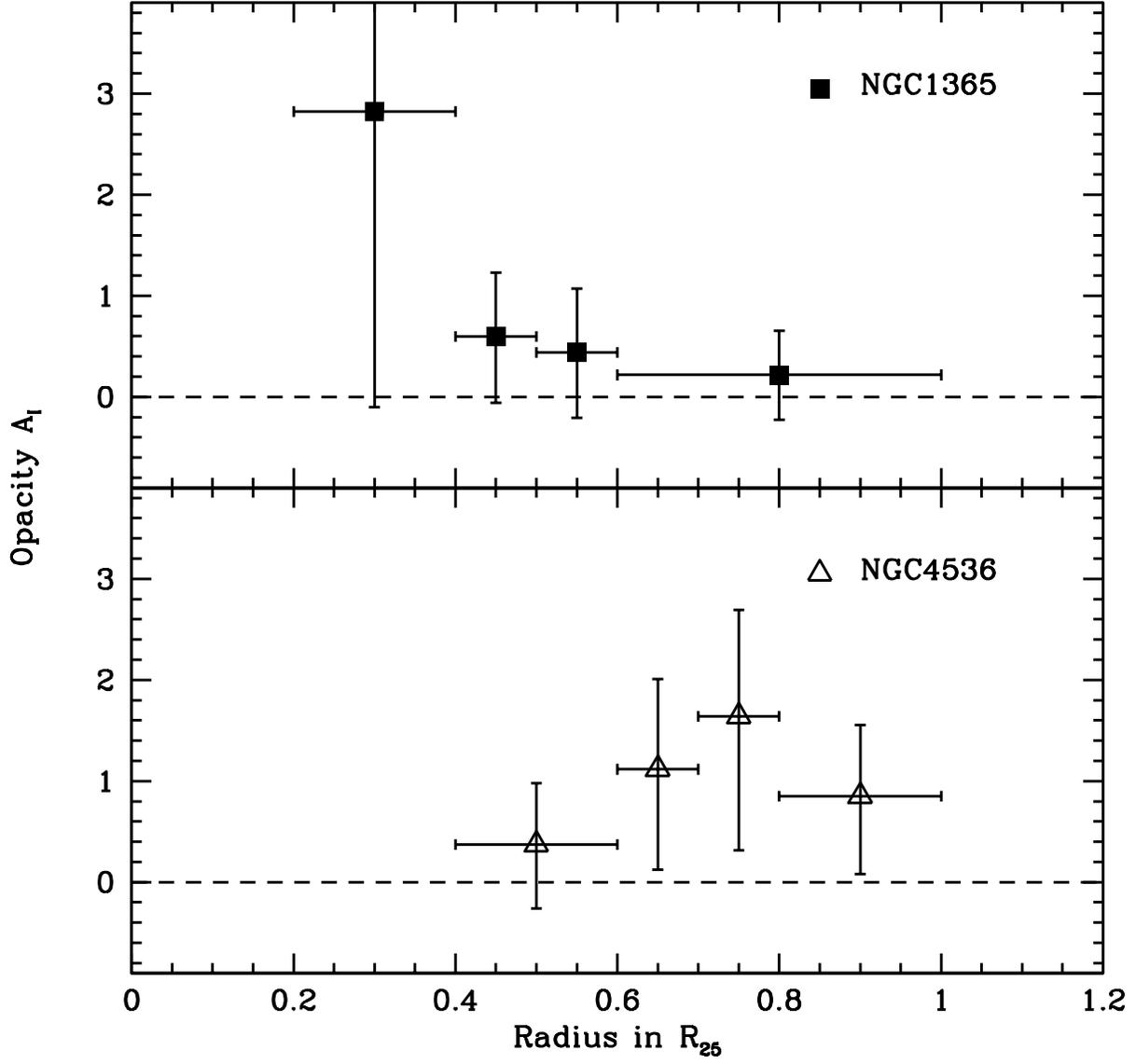}
\caption{The dust extinction plotted as a function of radius, expressed in $R_{25}$. 
The opacity of NGC 1365 (top) for the entire field, with all radii combined, is $A = 0.5_{-0.3}^{+0.3}$ and for NGC 4536 (bottom plot), the opacity of the entire field is, $A = 0.9_{-0.4}^{+0.4}$. }
\end{figure}
\clearpage

\begin{deluxetable}{l l l}
\tablewidth{0pt}
\tablecaption{Source Extractor Input Parameters}
\tablehead{\colhead{Parameter}      & \colhead{Value}          & \colhead{Comments} \\  }
\startdata 
PIXEL\_SCALE			& 0.05			& Scale in arcsec after drizzling\\
SEEING\_FWHM		& 0.17			& FWHM of the HST PSF 	\\
BACK\_SIZE			& 32				& Background estimation anulus.	\\
BACK\_FILTERSIZE		& 3 				& Background estimation smoothing factor.\\
BACKPHOTO\_TYPE	& LOCAL			& Photometric background.\\
BACKPHOTO\_THICK	& 32				& Photometric background anulus.\\
DETECT\_MINAREA	& 10 				& Minimum number of pixels in object.\\
FILTER				& Y				& Smooth before detection?\\
FILTER\_NAME		& gauss\_4.0\_7x7.conv 	& Smoothing kernel, gaussian with 4 pixel FWHM.\\
DEBLEND\_NTHRESH	& 32				& Number of deblending thresholds.\\
DEBLEND\_MINCONT	& 0.001			& Deblending minimum contrast.\\
CLEAN				& Y				& Remove bright object artifacts?\\
CLEAN\_PARAM		& 1.5     			& Moffat profile $\beta$ used for cleaning.\\
PHOT\_APERTURES	& 3,5,11,21,31		& Fixed aperture diameters.\\
GAIN				& 7.0				& Gain of the Wide Field CCD\\
\enddata
\end{deluxetable}

\clearpage
\newpage


\begin{deluxetable}{l l l l l r}
\tablewidth{0pt}
\tablecaption{Source Extractor Intrinsic Output Parameters}
\tablehead{\colhead{Name}      & \colhead{Description}          & \colhead{Unit}  & \colhead{Used}& \colhead{Comments}  \\ }
\startdata
A\_IMAGE         		& major axis 							& pixel		& * &	\\
B\_IMAGE         		& minor axis 							& pixel		&  & a	\\
ELLIPTICITY     	& 1 - B\_IMAGE/A\_IMAGE				& \nodata		& * &	\\
FWHM\_WORLD      & FWHM assuming a gaussian core 			& deg		& * &	\\
FLUX\_RADIUS     	& Fraction-of-light radii					& pixel		& * & b	\\
ISOAREA\_IMAGE   & Isophotal area above analysis threshold	& pixel$^2$	& * &	\\
CLASS\_STAR      	& S/G classifier output					& \nodata		& * & c	\\
MAG\_ISO         		& Isophotal magnitude					& mag		&  & d	\\
MAG\_AUTO        	& Kron-like elliptical aperture magnitude		& mag		&  &	\\
MU\_MAX          	& Peak surface brightness above background	& mag $/$ arcsec$^{2}$	& * & e	\\
MAG\_APER        	& Fixed aperture magnitude vector			& mag		& * & f	\\
\enddata

\tablenotetext{a}{B\_IMAGE was not used in the calculation of the galaxy score. The information is already contained in A\_IMAGE and ELLIPTICITY.}

\tablenotetext{b}{FLUX\_RADIUS is the radius in pixels containing a given percentage of the flux. $R_{eff}$ would be the FLUX\_RADIUS with 50\% of the light.}

\tablenotetext{c}{CLASS\_STAR is the SE output of a neural network
classification based on the relative areas of nine isophotes in each
object. It is only reliable for bright objects and becomes a random
value between 0 and 1 for fainter ones (\cite{SE}).}

\tablenotetext{d}{MAG\_ISO, the total flux of all the pixels above the
detection threshold. If the same pixels are selected in the other
filter by using dual image mode, the resulting color is more
indicative of the total object.}

\tablenotetext{e}{The ratio of MU\_MAX over MAG\_BEST (SE's choice between MAG\_ISO and MAG\_AUTO depending on crowding) provides a additional concentration index.}

\tablenotetext{f}{MAG\_APER, the flux within the specified apertures
(PHOT\_APERTURES). The fluxes from the V and I catalogs are a color indicator. For the colors we use an aperture with a diameter of  3 and 5 pixels (0\farcs15 and 0\farcs25 respectively). This choice of small diameters was done to obtain a conservative color estimate with minimal contamination from neighbouring objects in crowded fields.}
\end{deluxetable}

\clearpage
\newpage

\begin{deluxetable}{l l l l l  r}
\tablewidth{0pt}
\tablecaption{Source Extractor Output Parameters we have added }
\tablehead{\colhead{Name}      & \colhead{Description}          & \colhead{Used}  & \colhead{Unit} & \colhead{comments}  \\ }
\startdata
CONCENTRATION   	& Abraham concentration parameter			& * & \nodata	& a	\\
CONTRAST        		& Abraham contrast parameter 				& * & \nodata	& b	\\
SQR\_ASYMMETRY   	& Point-asymmetry index (difference squared)		&   & \nodata	& c	\\
ASYMMETRY       		& Point-asymmetry index (absolute difference)		& * & \nodata	& d	\\
MAJOR\_AXIS\_ASYM 	& Major axis asymmetry index					&  & \nodata	& e	\\
MINOR\_AXIS\_ASYM 	& Minor axis asymmetry index					&  & \nodata	& e	\\
MOFFAT          			& Computed Moffat magnitude					&  & mag	& f	\\
MOFFAT\_RMS      		& Ratio RMS deviation to computed Moffat flux.	&  & \nodata	& f	\\
MOFFAT\_RES      		& Ratio absolute residue to computed Moffat flux.	&  & \nodata	& f	\\
\enddata 

\tablenotetext{a}{CONCENTRATION is the fraction of light in the central
30\% of the objects area, measured in an ellipse aligned with the
object and having the same axis ratio. It is described in detail in
\cite{Abraham96} Adapted from code kindly provided by Dr. I. Smail. }

\tablenotetext{b}{CONTRAST is the fraction of object's flux in the
brightest 30\% of the total number of pixels. Also from
\cite{Abraham96} and courtesy of Dr. I. Smail.}

\tablenotetext{c}{SQR\_ASYMMETRY $ = \sum_i { (I_i - I_j )^2 \over I_i
+ I_j}$, where $I_j$ is the counterpart of $I_i$, equidistant with
respect to the object's center and rotated over 180$^{\circ}$. Described in 
\cite{Conscelice97} and adapted for SE by the authors.}

\tablenotetext{d}{ASYMMETRY $ = \sum_i {| I_i - I_j | \over I_i
+ I_j}$, where $I_j$ is the counterpart of $I_i$, equidistant with
respect to the object's center and rotated over 180$^{\circ}$. Based
on the expression in \cite{Conscelice00a} and also incorporated into SE.}

\tablenotetext{e}{MAJOR\_AXIS\_ASYM and MINOR\_AXIS\_ASYM are as
ASYMMETRY, but the x,y position of $I_i$ is mirror of the x,y position of $I_j$ with respect to the major or
minor axes respectively.}

\tablenotetext{f}{MOFFAT parameters: 
SE computes a Moffat profile ( $I = {I_0 \over (1+\alpha r^2)^{\beta} }$ ) 
from the peak pixel value ($I_0$) and the detection threshold (a known 
value of intensity ($I$) at a known distance ($r$) from the object's center). 
We hoped that star clusters and foreground stars could be picked out 
of our catalogs using their similarity to a typical Moffat profile. 
However, confusion from blends prevented an
easy selection. These parameters were not used in the computation of the Galaxy score.}

\end{deluxetable}

\clearpage
\newpage
\begin{deluxetable}{l  l l l l c c c}
\tablewidth{0pt}
\tablecaption{HST Archive Data Examples}
\tablehead{\colhead{Galaxy}  	& \colhead{Type} 	& \colhead{Prop. ID.}& \colhead{Exp Time}  		& 					& \colhead{Distance.}	 & \colhead{$R_{25}$} & \colhead{Inclination} \\
						&	 			& & \colhead{$V_{F555W}$}	& \colhead{$I_{F814W}$}	& \colhead{(Mpc.)} 		 & (Kpc.) & (Deg.)\\
 & & & & & (1) & (2) & (3)\\						
						}
\startdata
NGC 1365        & SBb     		& 5972 & 66560.0 	& 16060.0 	& 17.95	& 27.3  	& 34\\
NGC 4536        & SAB(rs)bc 	& 5427 & 68000.0		& 20000.0		& 14.93   	& 16.5	& 63\\
\enddata
\tablenotetext{1}{All distances were taken from \cite{KeyProject} }
\tablenotetext{2}{The 25 B-mag. surface brightness radius from the RC3 catalog \cite{RC3}. }
\tablenotetext{3}{Derived from the reported axis ratio ({\it sup\_ba}) in the 2MASS Large Galaxy Atlas \citep{LGA}.}

\end{deluxetable}

\clearpage 
\newpage


\begin{deluxetable}{l r r r r }
\tablewidth{0pt}
\tablecaption{Visual Rejection percentages} 
\tablehead{\colhead{Galaxy}  	& \colhead{Total} 	& \colhead{Crowded} 	& \colhead{Arm} & \colhead{Inter-arm}  \\ }
\startdata
NGC4536			&	22	&	89		& 	50	&	16\\
NGC1365			&	35	&	100		&	56	&	19\\
NGC1365	 (A=2)	&	55	&	97		&	66	&	31\\
adopted rates		&	-	&	100		&	50	&	20\\
\enddata
\tablecomments{Average rejection percentages of added HDF objects in visual checks identical to those in the real fields, using all the zero extinction simulations. We use rejection fractions for the average synthetic field counts of 0.5 and 0.2 for arm and inter-arm regions respectively to correct the simualted numbers for the visual step on the real number of field galaxies.}
\end{deluxetable}


\begin{deluxetable}{l l l l}
\tablewidth{0pt}
\tablecaption{Extinctions for different regions in the fields} 
\tablehead{
\colhead{Region}  	& \colhead{$A_I$} 	& \colhead{$\Delta A_c$} 	& \colhead{$A_I \times cos(i)$} \\
\colhead{}  		& \colhead{(1)} 		& \colhead{(2)} 			& \colhead{} \\
}
\startdata

NGC1365		& 					&				& 					 \\
WFPC2    		& $ 0.5^{+0.3}_{-0.3} $ 	& $ (\pm 0.3)	$ 	& $ 0.4^{+0.3}_{-0.3} $ \\
			&					&				&					\\
Arm(II)  		& $ 3.9^{+2.6}_{-3.5} $ 	& $ (\pm 1.8)	$ 	& $ 3.2^{+2.2}_{-2.9} $ \\
Arm(IV)  		& $ -0.7^{+1.1}_{-1.1} $ 	& $ (\pm 0.8) $ 		& $ -0.6^{+0.9}_{-0.9} $ \\
Interarm(III)    	& $ 0.4^{+0.6}_{-0.6} $ 	& $ (\pm 0.4)	$ 	& $ 0.3^{+0.5}_{-0.5} $ \\
Outside(V)       	& $ 0.5^{+0.4}_{-0.4} $ 	& $ (\pm 0.3)	$ 	& $ 0.4^{+0.3}_{-0.3} $ \\
			&					&				&					\\
R(0.2-0.4)       	& $ 2.8^{+2.4}_{-2.9}	$ 	& $ (\pm 1.6)	$ 	& $ 2.3^{+2.0}_{-2.4} $ \\
R(0.4-0.5)       	& $ 0.6^{+0.6}_{-0.7}	$ 	& $ (\pm 0.4)	$ 	& $ 0.5^{+0.5}_{-0.5} $ \\
R(0.5-0.6)       	& $ 0.4^{+0.6}_{-0.7}	$ 	& $ (\pm 0.4)	$ 	& $ 0.3^{+0.5}_{-0.5} $ \\
R(0.6-1.0)       	& $ 0.2^{+0.4}_{-0.4}	$ 	& $ (\pm 0.3)	$ 	& $ 0.2^{+0.3}_{-0.3} $ \\
\hline
\hline
			&					&				&					\\
NGC4536		& 					&				& 				 \\
WFPC2    		& $ 0.9^{+0.4}_{-0.4} $ 	& $ (\pm 0.3) $ 		& $ 0.4^{+0.2}_{-0.2} $ \\
			&					&				&					\\
WF2,3    		& $ 0.9^{+0.7}_{-0.8} $ 	& $ (\pm 0.5) $ 		& $ 0.4^{+0.3}_{-0.4} $ \\
WF4      		& $ 0.9^{+0.6}_{-0.7} $ 	& $ (\pm 0.4) $ 		& $ 0.4^{+0.3}_{-0.4} $ \\
			&					&				&					\\
R(0.4-0.6)       	& $ 0.4^{+0.6}_{-0.6}	$ 	& $ (\pm 0.4) $ 		& $ 0.2^{+0.3}_{-0.3} $ \\
R(0.6-0.7)       	& $ 1.1^{+0.9}_{-1.0}	$ 	& $ (\pm 0.6) $ 		& $ 0.5^{+0.4}_{-0.5} $ \\
R(0.7-0.8)       	& $ 1.6^{+1.0}_{-1.3}	$ 	& $ (\pm 0.7) $ 		& $ 0.7^{+0.5}_{-0.6} $ \\
R(0.8-1.0)       	& $ 0.9^{+0.7}_{-0.8}	$ 	& $ (\pm 0.5) $ 		& $ 0.4^{+0.3}_{-0.4} $ \\

\enddata
\tablenotetext{1}{Opacities from $A_I = -2.5~ C ~ log_{10}(N/N_0)$, errors are the $1 \sigma$ uncertainties, including the clustering uncertainty in the number of galaxies from the science field.}
\tablenotetext{2}{The contribution to the total error in opacity ($\Delta A_I$) owing to galaxy clustering uncertainty in the background.}
\tablenotetext{3}{The opacity and errors, corrected for inclination.}
\tablecomments{Extinction measures in the different regions in the WFPC2 mosaics, uncorrected and corrected for the inclinations from Table 4. }

\end{deluxetable}
\end{document}